\documentstyle[12pt,aasms4]{article}		

\newcommand{\xigg} {  {\xi_{gg}	       }	}
\newcommand{\xiGG} {  {\xi_{GG}	       }	}

\newcommand{\mlim} {  {m_{lim}	       }	}
\newcommand{\kms}  {  {\rm km  \ s}^{-1}	}
\newcommand{\hMpc} {  h^{-1}{\rm Mpc}  		}

\newcommand{\Mpc} {  {\rm Mpc}		}

\newcommand{\ltsima} {$\; \buildrel < \over \sim \;$}
\newcommand{\gtsima} {$\; \buildrel > \over \sim \;$}
\newcommand{\simlt}  {\lower.5ex\hbox{\ltsima}}
\newcommand{\simgt}  {\lower.5ex\hbox{\gtsima}}

\received{}
\accepted{}
\journalid{}{}
\articleid{??}{??}
\slugcomment{Submitted to ApJ, May 1996, astro-ph/9606116}

\lefthead{Trasarti Battistoni, Invernizzi, \& Bonometto}
\righthead{Clustering of Groups and Galaxies in PPS}

\begin{document}

\title{Clustering of loose groups and galaxies \\
from the Perseus--Pisces Survey}
\author{
		Roberto Trasarti Battistoni\altaffilmark{1}, 
		Gianluca Invernizzi\altaffilmark{2}, \&
		Silvio A. Bonometto\altaffilmark{3}
}
\affil{
		Dipartimento di Fisica dell'Universit\`a,  
		Via Celoria 16, 20133 Milano, Italy
}

\altaffiltext{1}{e-mail: trasarti@astmiu.uni.mi.astro.it}
\altaffiltext{2}{e-mail: laurbonom@astmiu.uni.mi.astro.it.}
\altaffiltext{3}{e-mail: bonometto@mi.infn.it.}

\begin{abstract}

We investigate the clustering properties of loose groups in the Perseus--Pisces
redshift Survey. Previous analyses based on CfA and SSRS surveys led to
apparently contradictory results. We investigate the source of such
discrepancies, finding satisfactory explanations for them. Furthermore, we find
a definite signal of group clustering, whose amplitude $A_G$ exceeds the
amplitude $A_g$ of galaxy clustering ($A_G=14.5^{+3.8}_{-3.0}$,
$A_g=7.42^{+0.20}_{-0.19}$ for the most significant case; distances are
measured in $\hMpc$). Groups are identified with the adaptive
Friends--Of--Friends (FOF) algorithms HG (Huchra \& Geller 1982) and NW
(Nolthenius \& White 1987), systematically varying all search parameters.
Correlation strenght is especially sensitive to the sky--link $D_L$ (increasing
for stricter normalization $D_0$), and to the (depth $\mlim$ of the) galaxy
data. It is only moderately dependent on the galaxy luminosity function
$\phi(L)$, while it is almost insensitive to the redshift--link $V_L$
(both to the normalization $V_0$ and to the scaling recipes HG or NW).

\end{abstract}

\section{Introduction}
\label{sez:intro}

Galaxy, group and cluster distributions probe matter clustering in the
Universe, not only over different scales, but also for different 
density contrasts. However, while galaxy and cluster clustering have 
been widely inspected, a measurement of group clustering meets
several conceptual and technical difficulties and it is not
surprising that its results are controversial and partially
contradictory.

In this note we report the result of an analysis of clustering properties of
loose groups in the Perseus--Pisces redshift Survey (hereafter PPS; see
Giovanelli, Haynes, \& Chincarini 1986; Haynes et al. 1988; Giovanelli \&
Haynes 1989, 1991, 1993). Through such analysis we believe that the reasons of
previous discrepant results become clear. It is also worth soon mentioning that
our error analysis, based on bootstrap criteria, detects a precise signal of
clustering for loose groups above statistical noise. 

As is known, the 2--point functions of galaxies and clusters are
consistent with the power laws
\begin{equation}
\label{eq:a_gc}
\xi_{gg} = A_g r^{-\gamma}~,
~~~~
\xi_{cc} = A_c r^{-\gamma}~,
\end{equation}
characterized by the same exponent $\gamma \simeq 1.8$, but by
widely different amplitudes $A_g$ and $A_c$. The detection of
such difference (Bahcall $\&$ Soneira 1983, Klypin $\&$ Kopylov 1983)
led Kaiser (1984) and Politzer $\&$ Wise (1984) to suggest
the mechanism of biased galaxy formation.

Results are far less clear for galaxy groups. Jing $\&$ Zhang (1988,
hereafter JZ88) and Maia $\&$ daCosta (1990, hereafter MdC90) claimed
that the 2--point function for groups is still consistent with a
power law $\xi_{GG} (r) = A_G r^{-\tilde \gamma}$ with ${\tilde \gamma }= 1.8$
and $A_G = A_g/d$ with $d \simeq 2$--$2.5\, $. On the contrary,
Ramella, Geller , \& Huchra (1990, hereafter RGH90) found
$\xi_{GG} (r) \simeq  \xi_{gg} (r)$ and, although their
analysis cannot reject a value ${\tilde \gamma} = 1.8$,
the preferred value ranges around 1$\, $. According to RGH90,
the main contribution to $\xi_{GG}$ comes from the the
2--point function $\xi_{mm}$ of galaxies members of groups.

Recently Frederic (1995a\&b, hereafter F95)
determined $\xi(r)$ for haloes and halo groups in
CDM simulations by Gelb (1992). 
He found groups to be significantly more correlated than single halos,
and interpreted this as contrasting with 
RGH90's results for galaxies and galaxy groups (but he 
also showed that the correlation strength depends on the 
prescription adopted for halo identification and illumination).

In all the above mentioned studies, groups were identified with the
adaptive Friends--Of--Friends algorithms of Huchra \& Geller 
(1982; HG82 hereafter) or Nolthenius \& White (1987; hereafter NW87).
Such algorithms require several input parameters. 
Some (the galaxy luminosity function $\phi(L)$ and the magnitude limit $\mlim$)
are set by the data themselves. 
Others 
(the ``sky--link'' $D_0$ and the ``redshift--link'' $V_0$)
must be decided by the user:
$D_0$ can be related to the normalization $\phi_*$ of $\phi(L)$ (NW87),
while the choice of $V_0$ is more complex
(HG82; NW87; Ramella, Geller, \& Huchra 1989, RGH89 hereafter).
As already pointed out by NW87, confirmed by RGH89, and stressed by Nolthenius,
Klypin, \& Primack (1994, 1995; hereafter NKP94\&95), a delicate point in group
analysis is the {\sl sensitivity} of the results to the details of the adopted
algorithm and/or data set. 
Also forgetting possible intrinsical differences
among the galaxy samples 
where groups were drawn from,
the different search parameters of the algorithm
used to identify galaxy groups 
could  be at the origin of the above mentioned discrepancies.
However, as we shall see below, this is
actually more relevant for internal than for clustering properties
(F95; Trasarti Battistoni 1995, 1996 -- TB96 hereafter).
Another problem is the high noise in the determination of $\xiGG$,
due to the limited extension of the group catalogs previously studied.

Loose groups in PPS were systematically identified and analyzed in TB96,
who concentrated mainly on internal properties and 
their dependence on the adopted algorithm and/or data sample.
Differences between data samples are small but detectable, and 
the effect of the magnitude limit $\mlim$ is to be properly taken into account.
Note that PPS is wider than the CfA2 Slices
(de Lapparent, Geller, \& Huchra 1986, 1988, 1989 -- dLGH86/88/89 hereafter;
Geller \& Huchra 1989; Huchra et al. 1990, Huchra , Geller, \& Corwin 1995)
used by RGH89--90 and F95, and is spatially disconnected
from them
as it lies in a different galactic hemisphere.
It is also deeper than the redshift surveys
CfA1 (Davis \& Huchra 1982, Davis et al. 1982, Huchra et al. 1983) and 
SSRS1 (da Costa et al. 1988), 
where groups identified by 
Geller \& Huchra (1983; hereafter GH83) and
Maia, daCosta, \& Latham (1989; hereafter MdCL89) 
were used by JZ88 and MdC90, respectively.
In fact, the number of groups in PPS is $\sim 180$-$200$,
while it is $\simlt 100$-$150$ in the other samples,
and this helps to reduce the above mentioned statistical noise.

Internal properties of groups have been used to constrain
cosmological models and, in particular, the
dark matter composition (NKP94\&95). Also group clustering has been
suggested as a test for  cosmological models, 
both on analytical bases (e.g., Kashlinsky 1987), 
or through the comparison with numerical N--body simulations (F95).
In the latter case, the key point is that
galaxy groups can be identified automatically and {\sl exactly in the same way} 
both from galaxy catalogs and from large 
($R \simeq 100 \ \hMpc$) N--body simulations
(NW87; Moore, Frenk, \& White 1993; NKP94\&95; F95).
Although such groups are basically expected to be physical objects
this is no longer the basic requirement to have
an effective comparison. Once groups are suitably defined, then
 properties
are compared to find out which simulation best matches the observations.

There is a precise physical reason
which favours loose groups over
single galaxies (and compact groups) or rich clusters
as a test of cosmological models.
At intermediate separations ($\simeq 10 \ \hMpc$),
mass scales ($M \simeq 10^{13} M_\odot$),
and density contrast ($\delta n / n \simgt 10$--$100$)
typical of galaxy groups,
gravitational evolution is still in the mildly non--linear regime.
Therefore,
LSS keeps memory of the shape of the post--recombination power spectrum $P(k)$.
At larger scales linearity keeps the LSS signal at a level too low
in respect to the noise,
so $P(k)$ is not easily detectable,
and the limited extension (in volume and number of objects)  of 
available observational samples is often a problem. 
At smaller scales stronger non--linear and non--gravitational effects 
complicate everything.

Moreover, the most widely studied observational samples of rich clusters 
(Abell 1958; Abell, Corwin, Olowin 1989) suffer of various biases
(Sutherland 1988; see Borgani 1995 for a review),
mainly due to partially subjective criteria used in their compilation.
Compact groups and rich clusters
were recently identified from observational samples 
also employing objective and automatic procedures 
(Prandoni, Iovino, \& MacGillivray  1994;
Nichol et al. 1992; Dalton et al. 1992; Nichol, Briel, \& Henry 1994).
However such procedures
are difficult to reproduce on N--body simulations.
This is due to a combined need of 
high resolution (to ease object identification),
large sample volume (to have a statistically meaningful number of objects),
and computational speed 
(to reach a statistically meaningful number of 
independent realizations of the same theoretical model).
In the case of clusters, the latter two difficulties can be circumvented by
using a combination of numerical and analytical approaches based on the
Zel'dovich approximation (e.g., Sahni \& Coles 1995; Borgani et al. 1995), but
the identification of observational--like clusters is still not an easy task. 

The plan of the paper is the following. 
In Section \ref{sez:data} we describe the galaxy data and the group catalogs,
while Section \ref{sez:xi} describes the estimation of clustering properties.
Results are presented and discussed in Section \ref{sez:resdis}.
We summarize our conclusions in
in Section \ref{sez:conclu}.

\section{Galaxy Data and Group Catalogs}
\label{sez:data}

The PPS database was compiled by Giovanelli \& Haynes in the last decade
(see Giovanelli \& Haynes 1991, 1993; Wegner, Giovanelli, \& Haynes 1993,
and the references therein). 
The full redshift survey is magnitude--limited down to $m_Z \leq 15.7$,
and now it covers the whole region 
$-2^h.00 \le \alpha \le +4^h.00$ and $ 0^o    \le \delta \le 50^o   $.
As in TB96, we restricted to the region 
$-1^h.50 \le \alpha \le +3^h.00$ and $ 0^o    \le \delta \le 40^o$,
to avoid regions of high interstellar extinction.
Magnitudes are anyway corrected as in Burstein \& Heiles (1978), and
redshifts are corrected for galactic rotation
and Local Group motion as in Yahil, Sandage, \& Tamman (1977).
The two subsamples PPS1 and PPS2 
(shown in Fig.~1, top panels) 
are magnitude--limited to $m_Z \leq 14.5$ and $15.5$ respectively, 
in analogy with CfA1 and CfA2.
This makes our comparison of data sample as clean as possible.
In fact, the selection criteria of SSRS
makes it qualitatively different from CfA or PPS.
On the other hand, the full CfA1 (North+South)
and the first two CfA2 Slices in the North
are neither fully disjoint (as PPS2 and CfA2 Slices) 
nor one a subset of the other (as PPS1 of PPS2).
The shape of the survey is also important.
It is more difficult to identify groups in the proximity of the edges,
so wide angle surveys are favoured over thin slices or pencil beams.
To summarize, PPS1 (PPS2) covers a solid angle $\omega=0.76 \ sr$,
and consists of
769 (3030)  galaxies 
with magnitude $ m_z \le 14.5  \ (15.5)$
and  redshift  $ cz  \le 17000 \ (27000) \ \kms$.
For comparison, 
the previously analyzed subsamples of
CfA1 N+S, SSRS1, and CfA2 Slices, are characterized respectively by:
$\omega=1.83+0.83$, $1.75$, $0.42$ $sr$, 
$m_Z \leq 14.5$, $\approx 14.5$ (apparent--diameter--limited), $15.5$,
$N_g=1534$, $1845+556$, $1766$.

The characteristics of all group catalogs are listed in Table 1.
The two significative cases TB96$_1$ and TB96$_2$ 
(in PPS1 and PPS2, respectively) are also shown in Fig.~1 (bottom panels).
Groups are identified with the friends--of--friends algorithms
described in TB96, both HG--like (HG82) and NW--like (NW87).
Briefly, two galaxies closer then some specified 
transverse separation $D_L$ and radial separation $V_L$ in redshift space
are friends of each other. 
Friendship is transitive, and a galaxy group is an isolated set of friends.
The two links are normalized by $D_0$ and $V_0$ at a given fiducial redshift
(here $cz_0=1000\ \kms$), and are then scaled--up with $cz$, using
the  selection function
\begin{equation}
\label{eq:psi}
\psi(cz; \mlim) = {
\int_{-\infty}^{M_{lim}} dM \phi(M) \over
\int_{-\infty}^{M_{fai}} dM \phi(M)
}~.
\end{equation}
Here $\mlim$ is the apparent--magnitude limit of the sample, 
$\phi(M)$ is the galaxy luminosity function,
$M_{fai}$ is the faintest absolute magnitude in the sample, while
the dependence on $m_{lim}$ and $cz$ arises through 
\begin{equation}
\label{eq:m_lim}
M_{lim} = m_{lim} - 25 - 5\log_{10}(cz/H_0)~.
\end{equation}

To scale up the links, 
the original HG prescription (based on simple arguments, Monte--Carlo tested)
gives 
$D_L \propto V_L\propto\psi(cz)^{-1/3}$,
while the NW recipes (based and tested on N--body simulations)
takes $D_L(cz) \propto (cz)^{-1/3} \psi(cz)^{-1/2}$,
and $V_L = V_0 + b \cdot (cz - cz_0)$, 
where $cz$ is the mean redshift of the pair of galaxies considered,
and $b=0.030$ is a suitable constant
(for a detailed discussion of the reasons behind such different choices, 
see TB96 and the references therein).
The value of $D_0$ corresponds to an effective density threshold
(in redshift space), given by
\begin{equation}
\label{eq:dn_n}
1+{\delta n \over n} = 
\bigg[{4\pi \over 3} D_0^3 \int_{-\infty}^{M_{lim}} dM \phi(M)\bigg]^{-1}
\end{equation}

We adopt the galaxy luminosity function in the Schechter (1976) form
determined from PPS2 by TB96
($\alpha=-1.15$, $M_*=-19.30$; here we take $\phi_* = 0.020\,h^3{\Mpc}^{-3}$).
The number of groups in all catalogs is approximately 
$N_G=0.06 \cdot N_g \propto \omega 10^{\mlim}$,
i.e. $N_G \sim 200$ in PPS2 and $N_G \sim 50$ in PPS1.
For sake of comparison, 
$N_G =88+47$,$87$, and $128$ in JZ88, MdC90, and RGH90 respectively.
It is clear that the characteristics of each groups catalogue
are fixed, to some extent, by the choice of parameters in the identification 
algorithm.
Smaller values of the links $D_0$ and $V_0$ yield groups  
with higher density contrast and, within biased theories of galaxy formation,
this is expected to cause a stronger spatial correlation. As outlined in TB96,
a similar effect arises from the sample depth. 
This subtler point requires some explanation. It is clear that
the increase of the mean galaxy density, which occurs when the
magnitude limit passes from $\mlim'$ to $\mlim'' > \mlim'$, is
accounted for by a slower decrease of the selection function
with $cz$. In fact, eq.~\ref{eq:m_lim} shows that
$\psi(\mlim'';cz'') = \psi(\mlim';cz')$, provided that
$cz'' = cz' 10^{m''_{lim} - m'_{lim}}$; accordingly, 
everything which happened at $cz'$ is now moved to 
$cz''$, in the deeper sample. Everything is therefore scaled, 
apart of $cz_0$, the redshift value where the sky--link $D_L$
is normalized.

Normalizing at the {\sl same} location $cz_0 = 1000 \ \kms$
for {\sl different} $\mlim$'s yields 
different density contrasts $\delta n / n$ for the same value $D_0$.
Groups identified with the same values of $D_0$ and $V_0$
are then expected to be more correlated in PPS1 then in PPS2.

An analogous effect is expected for the redshift--link $V_L$,
though different for the HG and NW algorithms. In fact, RGH90 suggested that
the different values JZ88 and MdC90 adopted for $V_0$ could account for the
discrepancies from their results, but they did not take into account the
stronger effect presumably arising from the different $\mlim$.
(The effects of changing $\mlim$, $D_0$, and $V_0$, are shown in Fig.~5;
a much more detailed discussion will be provided in Sect.~\ref{sez:resdis}.) 

It is also importants to outline that giving the $D_0$ link is not immediately
equivalent to providing a threshold density contrast. According to
eq.~(\ref{eq:dn_n}), $\delta n / n \propto \phi_* D_0^3$, but $\phi_*$
variations by a factor of 2 can arise both from observational techniques and/or
local physical conditions. The other parameters determining the Schechter
function, adopted by RGH89 and F95, and worked out by dLGH88 ($\alpha=-1.2$,
$M_*=-19.15$) are almost consistent with those worked out from our sample. TB96
showed that, for similar changes ($\Delta \alpha = \pm 0.1$ and $\Delta M_* =
0.1$), the net effect on the internal properties of output groups is small.
Furthermore, the spatial distribution of groups is almost insensitive even 
to greater differences in the identification algorithm. 

As shown by dLGH88 and Martinez et al. (1993),
estimates of $\xi(r)$ are sensitive to the presence of large scale features
in the samples.
In the present analysis, we cut {\sl all} samples within $cz \leq 11000 \ \kms$.
The number of groups does not change in PPS1 (767 galaxies),
while it is reduced by $\sim 5$--$10 \%$ in PPS2 (2693 galaxies).
This cut--off ensures that 
we are always dealing with the same physical structures,
though differently sampled by different $\mlim$'s.
The same main LSS features are present in both galaxy samples,
and they are reflected on the spatial distribution of groups
(Fig.~1).

\section{Two--point functions for galaxies and groups: estimate and errors}
\label{sez:xi}

Let us now describe the procedure leading to the 2--point
function estimate. As usual, 
we center spherical shells of radius $r$ and width $\delta r$ 
on each observed object in the data ($D$: galaxy/group) sample. 
We then count neighbours in the data sample
and in a random control sample, spatially uniform  
but with the same shape and radial selection function as the data.
To do so, 
$\tilde N_R$ random points are taken from a spatially uniform distribution 
with the same shape of the data sample,
but they are weighted down by the selection function $\psi(cz; \mlim)$.
In particular, 
the total number of random points $\tilde N_R$ is replaced by 
the total weight $N_R =  \Sigma_{k=1}^{\tilde N_R} w_k$,
where $w_k = \psi(cz_k; \mlim)$ is the weight of the $k-th$ random point.  
We then divide the number $DD(r;j)$ of (observed) neighbour objects within $r
\pm \delta r/2$ by the (weighted down) number $DR(r;j)$ of neighbour random
points, and then average over all $N_D$ centers. 
Our weighting scheme is analogous to that of the $\xi_{11}$ estimator 
discussed in dLGH88 (see also Martinez et al. 1993).
Altogether, this amounts to estimating $\xi(r)$ through
the following formula, that we use both for groups and galaxies:
\begin{equation}
\label{eq:xiest}
1+\xi(r)=  \langle 1+\xi(r;j) \rangle_{j=1,...N_D} =
{1\over N_D}\sum_{j=1}^{N_D} 
{DD(r;j) \over DR(r;j)} {N_R \over N_D}
\end{equation}
We adopt the same selection functions for galaxies and groups,
by using both in PPS1 and PPS2 the luminosity function $\phi_{STY}(M)$
evaluated in TB96 from PPS2 (in some cases, we use also $\phi(M)$ from dLGH88).
In fact, the
radial distribution of galaxies and groups substantially agree
over the relevant redshift range
(see Fig.~2).

Errorbars are computed with 10 bootstrap resampling of the data (Barrow,
Bhavsar, \& Sonoda 1984). Bootstrap errors are expected to overestimate the
true ensemble errors, in turn larger than formal Poisson errors. The ratio
between bootstrap and Poisson errors is expected to be $\sim 1.6$--$2.6$
(Martinez et al. 1993), with a slight dependence on $r$ and data depth. 
On the other hand, weighting data by $(1/$bootstrap errorbar$)^2$ in the
least--squares fit of $\xi(r)$ to $A r^{-\gamma}$, and simultaneously treating
$r$-bins as independent, tends to compensate the bootstrap overestimate,
yielding fair values for $A \pm \Delta A$ and $\gamma \pm \Delta\gamma$
(e.g., Ling, Frenk, \& Barrow 1986).

\section{Results and Discuss}
\label{sez:resdis}

Plots of 2--point correlation functions for groups are given in Figs.~3a and b.
For the sake of comparison, in each Figure, the galaxy 2--point functions are
from the same sample
also plotted. Here, groups are identified with $D_0=0.27 \ \hMpc$ and $V_0=350
\ \kms$, as in RGH89. Bootstrap error bars are given for all points.
A least--square bootstrap--weighted fit to $\xi(r)=A r^{-\gamma}$ is then
performed. The best fit $A$ and $\gamma$ and their errors are listed in 
Tables 2 and 3 for PPS1 and PPS2, respectively. 
For galaxies the fits can be extended from 1 to 31.7 $\hMpc$. For groups,
instead, the most reasonable distance interval is from 1.5 to 10 $\hMpc$. For
$r/\hMpc < 1.5$ anti--correlation due to the intrinsical size of groups is
expected to (and does) take over. The plots also show that, for $r/\hMpc > 10$
the signal is too noisy to be of any use. This distance interval is similar
to those used in previous group correlation analyses. We however performed
fits both for galaxies and groups in both kinds of distance intervals. 

From Fig.~3 we see that,
although (1) within 2--$\sigma$ bootstrap errorbars, $\xiGG \approx \xigg$,
(2) $\xiGG$ is mostly {\sl higher} than $\xigg$ by a factor $\simgt 2$. 
Accordingly, 
from Tables 2 and 3, we see that $A_G$ systematically exceeds $A_g$.
For the narrower and most reliable distance range, $A_G/A_g \sim 3$ for PPS1,
and $A_G/A_g $$\sim 2$ for PPS2. In this two--parameter fit, the
amplitudes are different at the $\sim$2.5--$\sigma$ level, but their difference
has the same sign anywhere, and can be suspected to be real. Also the slope
$\gamma$ of groups is however greater than the corresponding slope of galaxies.
Let us just remind that the range of $\gamma$'s found here is not unusual in
the redshift space and, for galaxies, corresponds to steeper $\tilde \gamma$'s
in the range $1.6$--$1.9$ in the real space 
(e.g., Gramann, Cen, \& Bahcall 1993).
The shift from redshift to real space is clearly expected to be weaker for
groups. It is not clear to which extent this can justify the steeper $\gamma$
found for groups; this effect exceeds 2--$\sigma$'s for PPS1 only, but is
present anywhere. 
With a comparable level of confidence we also
see that: (3) $A_G$ for PPS1 exceeds $A_G$ for PPS2 by more than 50$\%$, while,
in the same (narrower) distance range $A_g$ for PPS1 and PPS2 are almost
consistent within 1--$\sigma$. However, for galaxies, we can exploit the wider
distance range, and there we find a probable signal of luminosity segregation
at the 3--$\sigma$ level. 

The point (1) was also outlined by RGH90 in CfA2 Slices, while
the point (2) is in contrast with RGH90, MdC89, and JZ88. Let us
however remark that JZ88 and MdC89 used much larger $D_0$'s than us and RGH90.
Adopting the same links and $\phi(L)$ parameters as for JZ88's groups (GH83),
we obtain much lower and noisier $\xiGG$'s, both for PPS1 and PPS2. This is
shown in Fig.~4a, where, at variance with JZ88, error bars are also provided.
The discrepancy with RGH90, instead, is less pronounced. Besides of using a
different data set, RGH90 make also use of different $\phi(L)$ parameters. If
we make use of such parameters to detect groups in PPS2 we obtain a lower
$\xiGG$, nearly overlapping $\xigg$ (compare Fig.~3b and Fig.~4b).

An attractive interpretation of point (2) is a {\sl relative} segregation of
groups with respect to galaxies. A similar effect between halos and halo groups
in N--body simulations of flat unbiased CDM (Gelb 1992) was found  by F95, who
however showed that this result can depend on the identification scheme.
Although bearing in mind such reserves, we agree with F95 that RGH90's outputs
conflict with flat unbiased CDM models. However, this conflict is not evident
in our outputs. 

As already outlined, the obvious interpretation of the point (3) for galaxies
is luminosity segregation: 
as first shown by Davis et al. (1988) and convincingly demonstrated by
Park et al. (1994) 	
using complete {\sl volume--limited} samples, brighter galaxies are more
clustered than fainter one. 
This can be related to their greater luminosity either directly (e.g., Hamilton
1988; Hollosi \& Efstathiou 1988; Davis et al. 1988; Gramann \& Einasto 1992)
or through the luminosity--morphology connection (e.g., Iovino et al. 1993;
Hasegawa \& Umemura 1993). Such effect is also predicted by theoretical
scenarios (e.g., White et al. 1987; Bonometto \& Scaramella 1988;
Valls--Gabaud, Alimi, \& Blanchard 1989) and  a similar effect of {\sl mass}
segregation has been found in large N--body simulations (e.g., Campos et al.
1995). 

Independently of the interpretation, it is clear that cutting both PPS1 and
PPS2 within the {\sl same} limit $cz_{cut}=11000 \ \kms$, PPS1 contains all the
brighter galaxies of PPS2 because of the brighter $\mlim$. For these {\sl
incomplete} samples, limited both in $cz$ and in $m$, we expect $\xigg$ to be
greater for brighter $\mlim$. On the contrary, for {\sl complete}
apparent--magnitude limited samples ($cz_{cut}$ increasing with $\mlim$), we
expect $\xigg$ to be greater for fainter $\mlim$ (e.g., Hamilton 1988), as the
total number of intrinsically bright galaxies in such samples increases with
fainter $\mlim$ and dominates the sample. 

A possible interpretation of the point (3) for groups is that, in PPS1, we
selected higher density contrast groups than in PPS2. 
In fact, using the same $D_0$ for both PPS1 and PPS2, while the
mean inter galaxy separation is smaller for fainter $\mlim$,
can enhance the correlation in PPS1.
But this effect might not be enough.
In fact we checked that the change of $D_0$ in either PPS1 or PPS2 has smaller
effects than passing from PPS1 to PPS2. This is shown in Fig.~5a and b, where
we give $\xiGG$ for groups selected in 2 different ways both in PPS1 and PPS2.
Using $D_0 = 0.24 \hMpc$ ($D_0 = 0.30 \hMpc$) in PPS2 (PPS1) we have the same
$\delta n/n$  that $D_0 = 0.27 \hMpc$  gives in PPS1 (PPS2). The changes of
$\xiGG$ are however slight within either PPS1 or PPS2, in spite of the
change of $D_0$. 

We explore further the dependence of $\xiGG$ on $D_0$ in Fig.~5c.
As expected, $\xiGG$ increases with $\delta n/n$ within a given sample
(the Figure refers to PPS2).
In Fig.~5d, instead, we test the dependence on $V_0$ and discover that, at
variance with what could be expected, $\xiGG$ does not depend on $V_0$, in
spite of varying the redshift link over a greater range than $D_0$. As a
further check, we also used the NW scaling recipe, whose $V_L$ has a quite
different shape from the HG scaling ($D_L$'s are very similar). 
No effect on $\xiGG$ is however present. 
(This agrees with F95, and is consistent with TB96 who showed how 
for given self--consistent normalizations and different scaling recipes
the spatial distribution of HG--like and NW--like groups is very similar.)
It is difficult to comment on this point, which however seems to indicate that
varying $V_0$ is not so effective to change the density contrast and that the
redshift link is therefore less {\sl physical} than the sky link. In fact, TB96
showed how a restrictive $V_L$ affect groups more as a cut--off (discarding
high--velocity--dispersion systems) than as a systematic variation of the
properties of each single group. On the other hand, changing $D_L$ directly
affects $\delta n/n$ of the otput groups. From a physical point of view, this
might suggest that clustering of galaxy systems is more directly related to
their density contrast than to their velocity dispersion (mass?). 

Note that groups are on average brighter in PPS1 than in PPS2 (TB96).
A natural interpretation of point (3) could be 
luminosity (mass?) segregation among {\sl groups} themselves. 
But this could be a premature conclusion.
In fact, in order to compare group luminosity the 
{\sl observed} group luminosity $L_{obs}$
should be implemented by $L_{fai}$ for group members below $\mlim$.
But this correction depends on several assumptions 
(Mezzetti et al. 1985; 		
Gourgoulhon, Chamaraux, \& Fouqu\`e 1992; 	
Moore et al. 1993).		
The enhancement of $\xiGG$ in PPS1 could be just an effect of a higher
correlation of group members. In fact, groups are brighter in PPS1 than in PPS2
because their {\sl members} are. Henceforth, luminosity segregation of 
{\sl galaxies}, not of groups themselves. 

However, Fig. ~1 shows that the relatively nearby main ridge in PPS at
$cz\approx 5000 \ \kms$ is equally well sampled in both PPS1 and PPS2, while
the more sparsely populated region between $6000$ and $11000\ \kms$ in PPS2 is
completely devoid of groups in PPS1. From this point of view, groups {\sl are}
indeed more clustered in PPS1 than in PPS2, but simply because of
``cosmographical'' reasons, not because of their higher luminosity. Therefore,
an explanation of effect (3) could be an intrinsic difference in the data sets
PPS1 and PPS2. 

Previous analyses of group clustering (JZ88, MdC90, RGH90)
led to apparently contradictory results. 
The main source of discrepancy lies in the different search parameters of the
adaptive Friends--Of-Friends algorithms adopted by various authors. The greater
difference is due to the transverse normalization $D_0$.
Stricter $D_L$'s yield stronger correlations, thus explaining the discrepancy
among JZ88 on one side ($\xigg \approx 2.5 \xiGG$) and RGH90 on the other
($\xigg \approx \xiGG$).
Instead, contrary to simple expectation, no dependence on the radial link $V_L$
(as suggested by RGH90) is detected. 
A further reason of difference is the impact on group properties
of the different depth of the adopted samples. 
This effect (partially considered by RGH89 and MdCL89, but not by RGH90
and MdC90) 
requires a suitable handling of the ``parameter'' $\mlim$ in the
identification algorithm, and a self--consistent normalization of $D_L$. 
Simply adopting the same value of $D_0$ leads to lower density contrasts
in shallower samples, also contributing to the shallower $\xiGG$ of JZ88
and MdC90.
The residual difference among samples of different depth is nevertheless
more important.
This is the case for the samples PPS1 and PPS2, and
an analogous effect was pointed out by RGH89 for CfA1 Survey and CfA2 Slices.
A further reason of discrepancy between RGH90 and MdC90 could be 
the different physical nature of the CfA and SSRS samples.
As shown by MdCL89, groups in CfA1 and SSRS1 have similar physical properties
when {\sl different} values of the links are used. In fact,
the former survey contains a higher percentage of early--type galaxies,
in general more clustered than the late--type galaxies in the latter.

\section{Conclusions}
\label{sez:conclu}

The attention devoted by a number of authors to the clustering
of loose groups is widely justified. Galaxies and clusters
are both clustered, but their clustering parameters are different.
The discovery of such difference was the original motivation
to introduce the biased theory of galaxy formation. An independent
test of such theory as a whole and a further constraint on its
parameters arise from a sufficiently precise determination
of clustering parameters of groups. 

To pursue this program we need an unambiguous definition of
loose groups and a sample wide enough to test their properties.
In the literature there has been a complex debate on
group individuation, which is technically provided by suitable
sky link ($D_L$) and redshift link ($V_L$), in transverse
and radial direction, respectively. This is due to the need
of exploiting the whole information contained in apparent
magnitude limited samples, and overcoming the critical problem
of distortions due to the passage from the redshift to the
real space. 

This work shows that results on clustering parameters, obtained with different
group individuation recipes, are essentially consistent. This does not mean that
the values of the links do not matter. On the contrary, the clustering strength
is critically dependent on how restrictive the selection is. The point is that
$D_L$ has a stronger impact on clustering properties than $V_L$. Different
criteria (e.g., NW vs. HG) are essentially different as far as the radial link
$V_L$ is concerned, but the effect on $\xi(r)$ of fairly wide variations of
$V_L$ is neglegible, while smaller variations of $D_L$ are reflected on the
clustering of groups. 

To some extent, this is a natural outcome of working in {\sl redshift space}.
The transverse link $D_L$ is directly related to the spatial separation among
group members. In the radial direction, instead, peculiar velocities inside
groups yield ambiguous apparent separations. Relating $V_L$ to a value of
$\delta n /n$ is then not trivial and depends on the cosmological model (e.g.,
NW87). 

This problem would be present even if we could select a complete
volume--limited galaxy sample, and then identify groups therein as suggested by
Ramella et al. (1995a,b).	
(Selecting volume--limited subsamples of the group catalogs {\sl after}
group identification with the adaptive HG and NW algorithms
(Zabludoff et al. 1993a,b),	
is not the ideally correct solution, as groups could be already biased by the
magnitude--limited nature of the galaxy data source.) Unfortunately, to be
really efficient, this procedure requires a high number of measured redshifts.
$E.g.$, even adopting the {\sl generous} constant links $D_0=0.5 \ \hMpc$,
$V_0=350 \ \kms$, volume limiting PPS2 to $cz \leq 7900 \ \kms$ ($M \leq
-19.5$) as in Guzzo et al. (1991) and Bonometto et al. (1993) yields only $N_G
\approx 50$ groups, and it is not sure that this is enough. 

Resorting then to {\sl apparent--magnitude--limited} samples, two
further but connected source of complication arise.
First, the physical mixture of what we consider {\sl bright} and {\sl faint}
galaxies varies with redshift. 
Second, the mean separation among observed galaxies grows with $cz$,
and group identification is harder at higher redshifts.
To some extent, the traditional strategy of compensating the decrease of galaxy
density by scaling up $D_L$ and $V_L$, gives group properties doomed to be 
systematically dependent on $cz$ (and, to a lesser extent,
to the scaling recipe and galaxy luminosity function).

Previous analyses of group clustering (JZ88, MdC90, RGH90) in CfA and SSRS
surveys led to apparently contradictory results. In this paper, we have
investigated the source of such discrepancies, finding satisfactory
explanations for them. Together with that, we have found a signal of group
clustering, whose amplitude exceeds the amplitude of galaxy clustering. Such
excess is perhaps to be trusted more than what its formal probability ($\sim
2.5\, \sigma$'s) prescribes, as it persists through various density contrasts
and link recipes; it could be found mostly thanks to the wider extension of
PPS in respect to the sample used for previous analyses. 

\acknowledgments{We are grateful to R. Giovanelli and M. P. Haynes 
for discussions, and for kindly providing us with the PPS data.}

\clearpage
\begin{deluxetable}{crrrr}
\footnotesize
\tablecaption{FOF algorithms and group catalogs.} 
\tablewidth{0pt}
\tablehead{
\colhead{Group Catalog} 	&
\colhead{$D_0$ }						&
\colhead{$V_0$ }						&
\colhead{$\delta n/n$}						&
\colhead{Number of groups}			
}

\startdata
                &	& 	&   		&	 \nl
TB96$_{1,2}$	&0.27	&350	&160,110	& 48, 192\nl		
                &	& 	&   		&	 \nl
wide$_1$	&0.30	&350	&110, ...	& 48, ...\nl
strict$_2$	&0.24	&350	&..., 160	&..., 179\nl
high$_{1,2}$	&0.21	&350	&330, 240	& 40, 168\nl
low$_{1,2}$	&0.42	&350	& 40,  30	& 59, 205\nl
cold$_{1,2}$	&0.27	&150	&160, 110	& 43, 186\nl
hot$_{1,2}$	&0.27	&600	&160, 110	& 50, 201\nl
TB96NW$_{1,2}$	&0.27	&350	&160, 110	& 48, 162\nl
RGH89$_{1,2}$	&0.27	&350	&120,  80	& 49, 192\nl
RGH83$_{1,2}$	&0.27	&600	&120,  80	& 52, 205\nl
DGH83$_{1,2}$	&0.52	&600	& 16,  10	& 60, 185\nl
                &	& 	&   		&	 \nl
\enddata

\tablecomments{Indexes $1$, $2$ stand for the galaxy samples PPS1, PPS2
where groups are drawn from.
Scaling recipes are everywhere as in HG82, except for TB96NW
as in NW87. The reference algorithm TB96 is detailed as follows:
wide$_1$/strict$_2$  means wide/strict $D_0$ (in $\hMpc$),
so to match $\delta n /n$ of the
samples TB96$_2$, TB96$_1$ respectively (see text); high/low means high/low
density contrast $\delta n /n$, i.e. small/large $D_0$; cold/hot means low/high
velocity dispersion $\sigma_v$, i.e. small/large $V_0$ (in $\kms$).
The galaxy luminosity function is everywhere as in TB96, except for 
RGH89, RGH83, and DGH83}
\end{deluxetable}

\clearpage
\begin{deluxetable}{crrr}
\footnotesize
\tablecaption{A and $\gamma$ for groups and galaxies in PPS1.}
\label{tab:fit1}
\tablewidth{0pt}
\tablehead{
\colhead{Sample}			&
\colhead{$\gamma \pm \Delta \gamma$  }	&
\colhead{$A \pm \Delta A$}		&
\colhead{$r_0 \pm \Delta r_0$}
}

\startdata

{\bf PPS1} &$1.21\pm .04$&$10.54+.75-.70$&$7.03 +0.62-0.60$\nl
TB96$_1$  &$1.32\pm .16	$&$24.2 +6.4-5.0$&$11.2+4.0-3.8	$\nl
                 &  & & \nl
PPS1	  &$1.08\pm .06	$&$9.5 +0.9-0.8	$&$8.0 +1.2-1.1$\nl
{\bf TB96$_1$} &$1.39\pm .26 $&$30.2+15-9.8$&$11.6+6.6-5.9$\nl
wide$_1$  &$1.50\pm .29	$&$42  +23-15	$&$12.0+7.2-6.4$\nl
                 & & &  \nl
high$_1$  &$1.32\pm .32	$&$41  +25-16	$&$17+14-13$\nl
low$_1$	  &$1.20\pm .28	$&$11.3+5.4-3.6	$&$7.5 +4.6-4.0	$\nl
                 & & &  \nl
NW$_1$	  &$0.70\pm1.25	$&$10.0+5.3-3.5	$&$27+37-34	$\nl
cold$_1$  &$1.26\pm0.28	$&$26.4+13-8.5$&$13.4+9.3-8.5	$\nl
hot$_1$	  &$1.51\pm0.33	$&$44+30-18     $&$12.5+8.8-7.6	$\nl
                 & & &  \nl
RGH89$_1$ &$1.59\pm .28	$&$50	+27-18	$&$11.7+6.5-5.7 $\nl
RGH83$_1$ &$1.36\pm .25	$&$43	+20-14	$&$16.0+10-9.1$\nl
DGH83$_1$ &$0.67\pm .74	$&$2.0  +5.3-1.5$&$3.9+12-5.0	$\nl
                 & & & \nl

\enddata

\tablecomments{Two--parameters least--square fit 
of $\xi(r)=A r^{-\gamma}$ to galaxy and group data (within the same sample);
$1$-$\sigma$ bootstrap errors are provided.
In the first two lines the distance range 
$1.0 \leq hr/\Mpc \leq 31.7$ is considered.
In the other lines $1.5 \leq hr/\Mpc \leq 10.0$.
The wider interval is actually suitable to fit the 2--point
function for galaxies only (see text).
Different group catalogs are obtained from different grouping criteria
(see text and Table 1),
but the most significant outputs are the $1^{st}$ and $4^{th}$ lines
(marked in bold characters).} 

\end{deluxetable}

\clearpage
\begin{deluxetable}{crrr}
\footnotesize
\tablecaption{A and $\gamma$  for groups and galaxies in PPS2.}
\label{tab:fit2}
\tablewidth{0pt}
\tablehead{
\colhead{Sample}			&
\colhead{$\gamma \pm \Delta \gamma$   }	&
\colhead{$A \pm \Delta A$}		&
\colhead{$r_0 \pm \Delta r_0$}
}

\startdata
                 & & &  \nl
{\bf PPS2} &$1.26\pm .02$&$7.42	+.20-.19$&$4.92+.18-.17	$\nl
TB96$_2$ &$1.31	\pm .14	$&$12.5	+2.9-2.3$&$6.9 +1.8 -1.7$\nl
                 & & &  \nl
PPS2	 &$1.21	\pm .04	$&$7.65+.36-.34	$&$5.36+.35-.34	$\nl
{\bf TB96$_2$} &$1.36\pm .19$&$14.5+3.8-3.0$&$7.2+2.5-2.2$\nl
strict$_2$ &$1.37\pm .18$&$15.4+4.4-3.4 $&$7.3+2.4-2.2$\nl
                 & & & \nl
high$_2$ &$1.04	\pm .19	$&$9.2 +3.1-2.3	$&$8.6+4.4-4.0$\nl
low$_2$	 &$1.58	\pm .23	$&$9.5 +3.2-2.4 $&$4.1+1.2-1.1$\nl
                 & & & \nl
NW$_2$	 &$1.08	\pm .19	$&$9.4 +3.4-2.5	$&$7.9+3.9-3.5$\nl
cold$_2$ &$1.28	\pm .13	$&$14.0+2.5-2.1	$&$7.9+2.0-1.9$\nl
hot$_2$	 &$1.29	\pm .22	$&$13.5+5.6-4.0	$&$7.5+3.5-3.1$\nl
                 & & & \nl
RGH89$_2$ &$1.63\pm .23	$&$16.0+5.5-4.1	$&$5.5+1.7-1.6$\nl
RGH83$_2$ &$1.45\pm .24	$&$12.7+5.0-3.6	$&$5.8+2.3-2.0$\nl
DGH83$_2$ &$0.83\pm .38	$&$1.9 +1.6-0.9	$&$2.2+2.3-1.4$\nl
                 & & &  \nl

\enddata

\tablecomments{Same as for Table 2, but for galaxies and groups in PPS2. The
comparison of the two bold face lines ($1^{st}$ and $4^{th}$) yields one of the
main results of this paper.} 

\end{deluxetable}


\clearpage

\figcaption{Wedge diagrams ($\alpha$--$cz$) of galaxies and loose groups
in the Perseus--Pisces redshift Survey
(only the region with low galactic extinction
$-1.5^h \leq \alpha \leq +3.0^h$,
$0^o \leq \delta \leq + 40^o$; 
all samples are cut within  $cz_{cut} = 11000 \, \kms$).
Groups are identified with the adaptive Friends--Of--Friends algorithm
of Huchra \& Geller (1982) (for definitions, see text).
(a) Galaxy sample PPS1.
(b) Galaxy sample PPS2.
(c) Group sample TB96$_1$.
(d) Group sample TB96$_2$.
}

\figcaption{Ratio $dN_G/dN_g$ between 
the number $dN_G$ of (identified) groups and 
the number $dN_g$ of (observed) galaxies
in the galaxy sample PPS2.
Three typical cases 
(with different search parameters $D_0$, $V_0$, and $\phi(L)$ -- see Table 1)
are shown.
Within the relevant distance range $2000 \simlt cz/\kms \simlt 10000$,
the radial distribution of groups and galaxies are in good agreement --
i.e., groups and galaxies share the same radial selection function
($dN_G/dN_g \sim 1/15$ independently of $cz$).
}

\figcaption{Two--point correlation functions 
$\xi_{GG}(r)$ for loose groups (thick lines) and 
$\xi_{gg}(r)$ for galaxies (thin lines),
within the {\sl same} galaxy sample(s):
(a) PPS1;
(b) PPS2.
Errorbars are $\pm 1$--$\sigma$ bootstrap.
In both samples, $\xi_{GG} \simgt 2 \xi_{gg}$ systematically,
though $\xi_{GG} \approx \xi_{gg}$ within $2$-$\sigma$ errorbars.
Also, both $\xi_{gg}(r)$ and $\xi_{GG}(r)$ 
in PPS1 are slightly {\sl higher} than in PPS2.
} 

\figcaption{Comparison of $\xi_{GG}(r)$ in PPS2 
for the loose group sample TB96$_2$ 
and other samples identified as in previous analyses:
(a) As in JZ88  
($D_0=0.52$, $V_0=600$ as in GH83; $\phi(L)$ as in RGH89),
for both PPS1 and PPS2;
(b) As in RGH89/90 ($D_0=0.27$, $V_0=350$; $\phi(L)$ as in RGH89),
for PPS2.
(The parameters of TB96$_2$ are: 
$D_0=0.27$, $V_0=350$, and $\phi(L)$ as in TB96.) 
The high noise in DGH83$_1$ is partially due to the low number of groups 
($N_G=60$), 
but the same degree of noise is also present in DGH83$_2$ ($N_G=185$).
The errorbars of RGH89$_2$ are similar to (but sometimes larger than) 
those of TB96$_2$, though $N_G=192$ in both cases.
Compare the small effect of $\phi(L)$ 
with those due to $D_0$ and $\mlim$ (see Fig.~5).
}

\figcaption{Dependence of $\xi_{GG}(r)$ in PPS1/PPS2 on 
$\mlim$, $\delta n/n$, $D_0$, and $V_0$ 
(scaling recipe as in HG82, $\phi(L)$ from TB96 -- see text):
(a) TB96$_1$ with 
the ``standard'' 	$D_0=0.27$, and 
the ``renormalized'' 	$D_0=0.30$ (wide$_1$) 
yielding the same $\delta n/n$ of TB96$_2$;
(b) TB96$_2$ with 
the ``standard'' 	$D_0=0.27$, and 
the ``renormalized'' 	$D_0=0.24$ (strict$_2$) 
yielding the same $\delta n/n$ of TB96$_1$;
(c) TB96$_2$ with its high-- and low--density counterparts;
(d) TB96$_2$ with its high-- and low--velocity--dispersion counterparts
(hot$_2$ and cold$_2$, respectively). 
The analogous group sample TB96NW$_2$, 
where both links (normalized as for TB96$_2$) were scaled
with the recipe of NW87, is also shown.
Errorbars (omitted for clarity) are always similar to those in Fig.3.
Note how $\xi_{GG}$ is only moderately sensitive to variations of $D_L$, 
while is almost independent from $V_L$;
it is particularly sensitive to the ``parameter'' $\mlim$
(in fact, any intrinsic difference between the galaxy data sets PPS1 and PPS2).
}

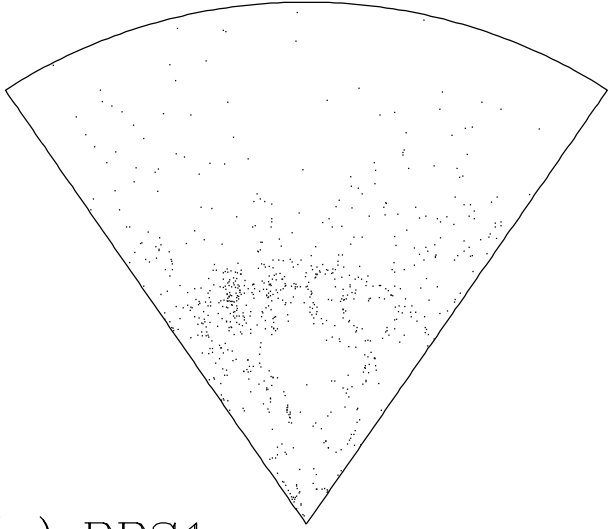
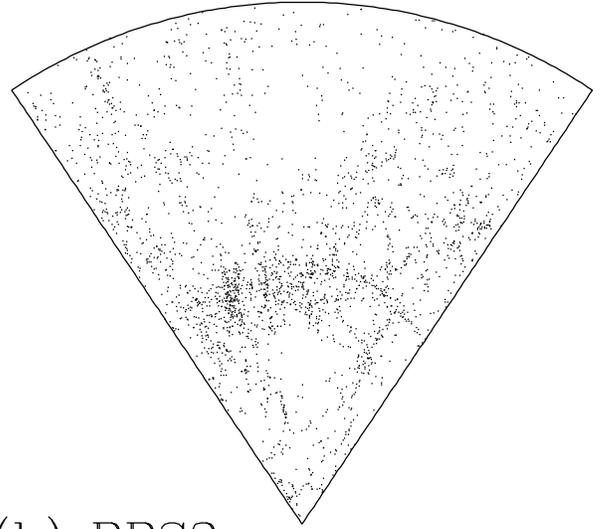
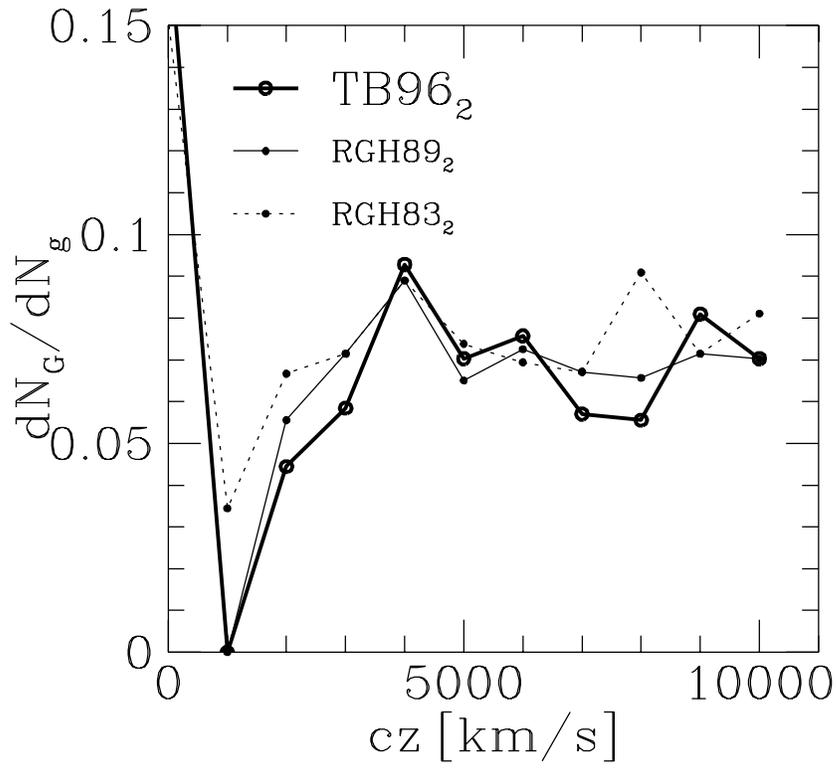
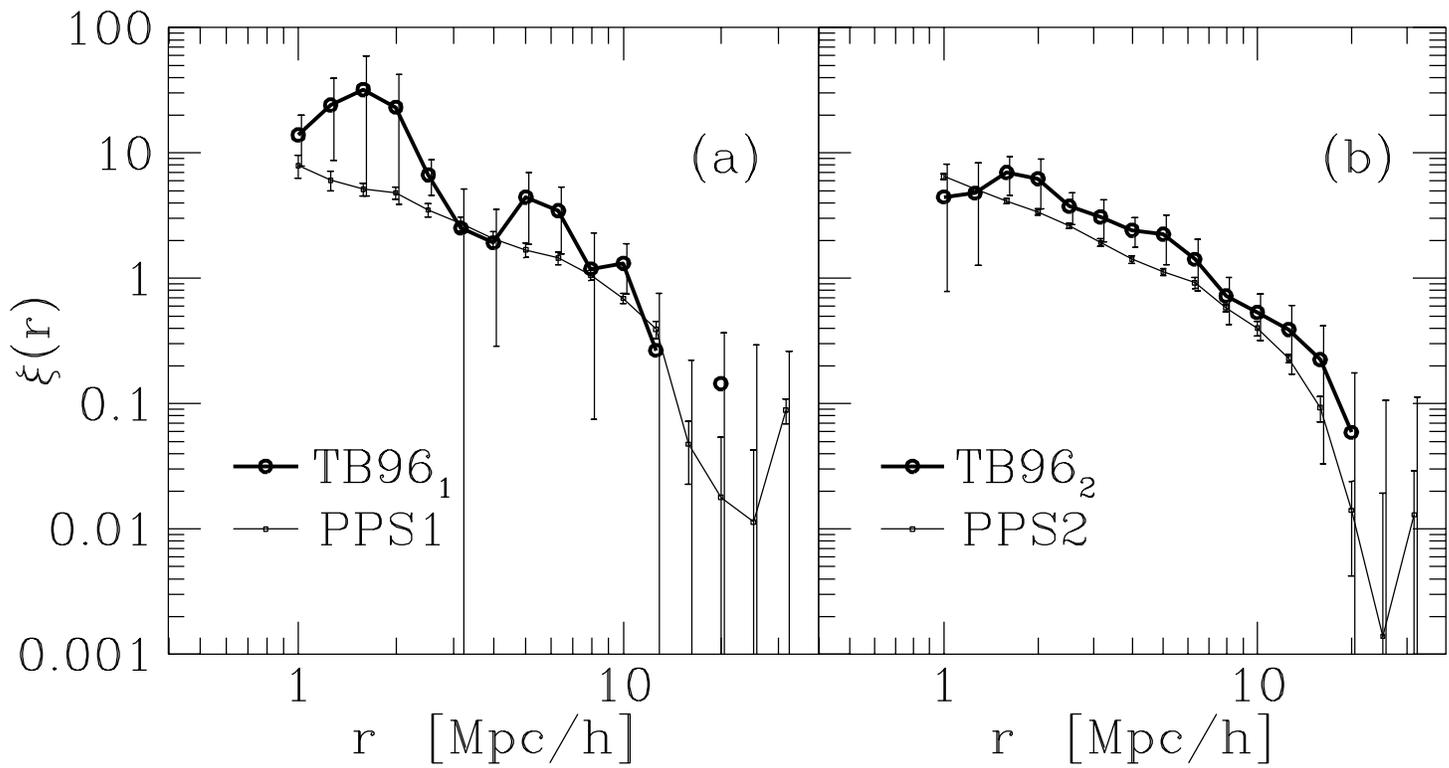
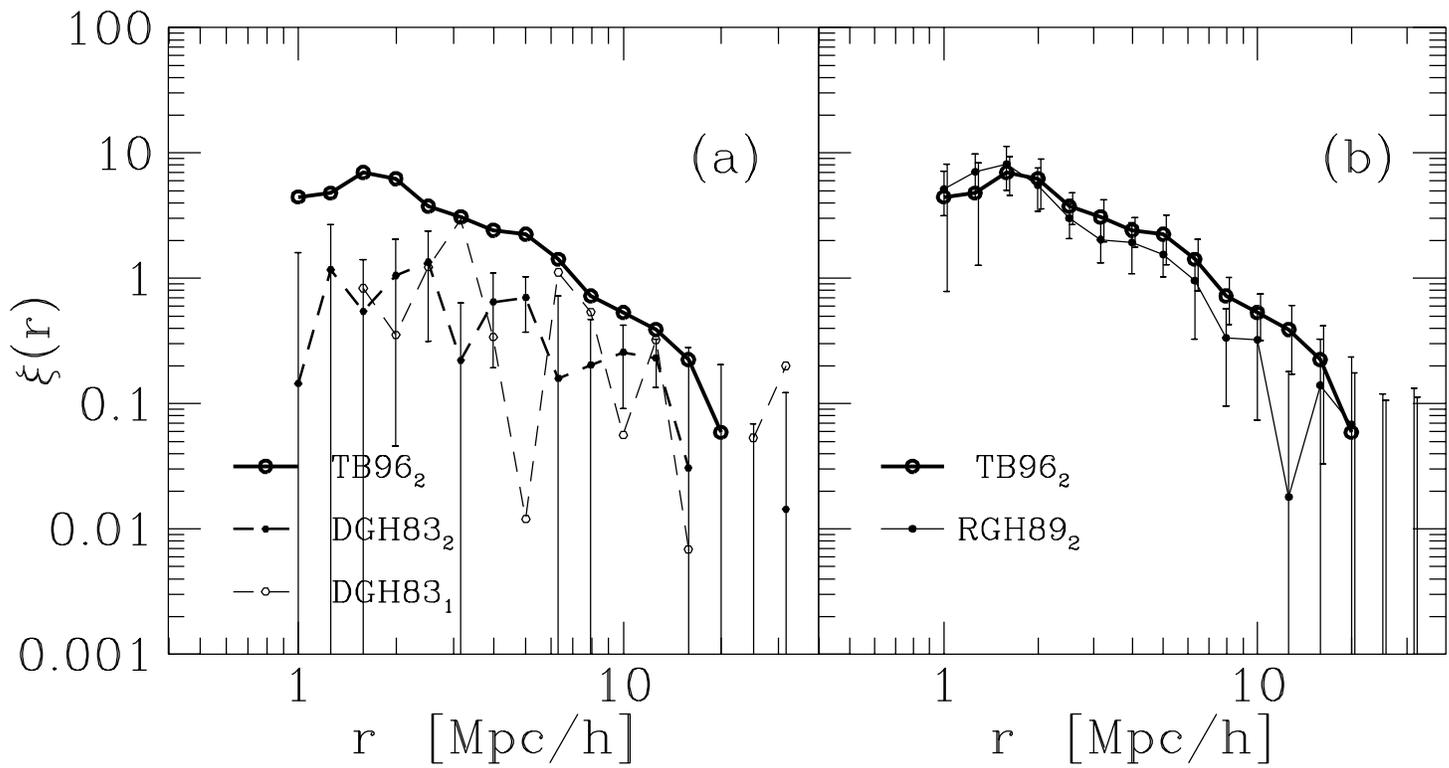
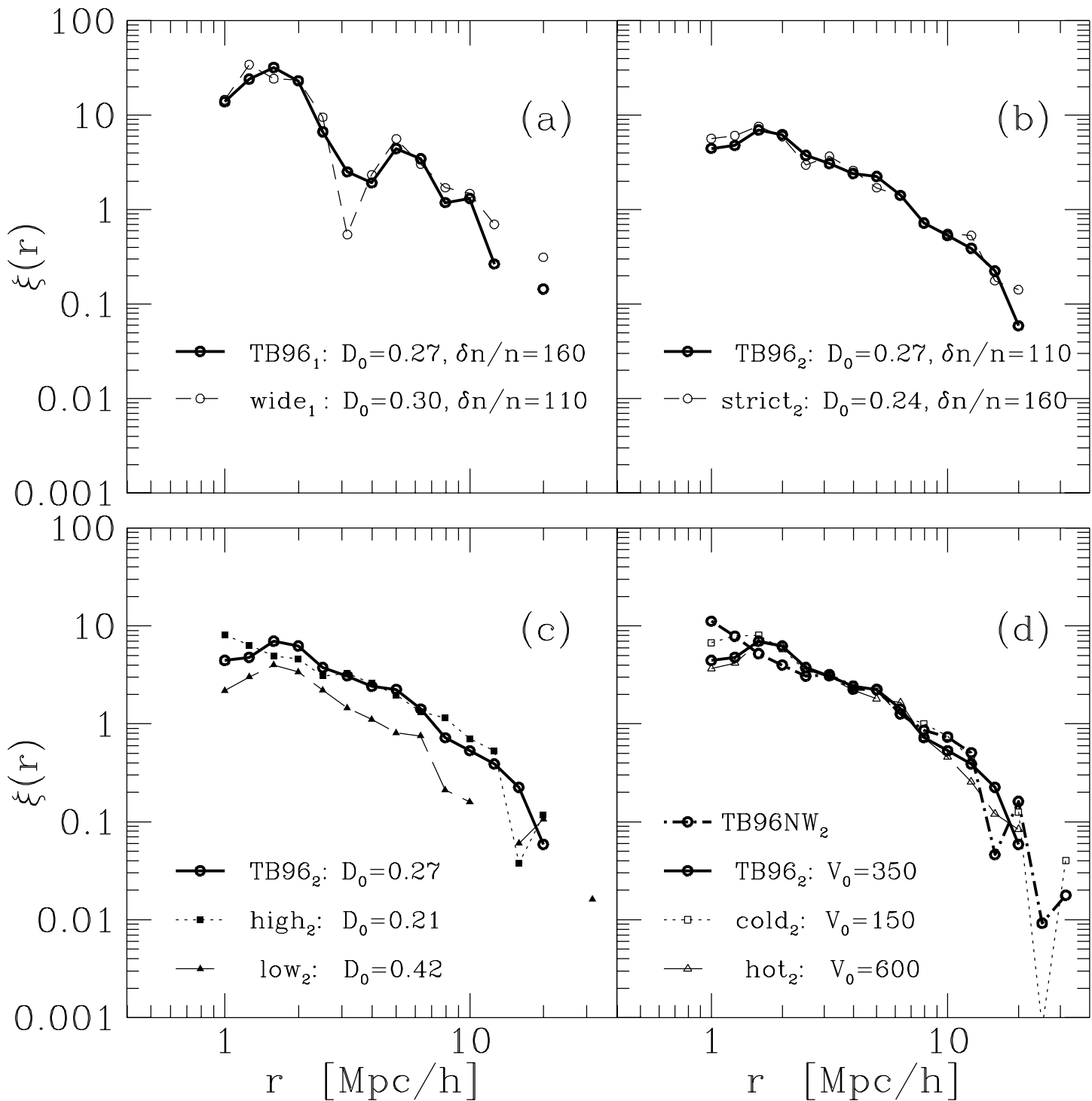
\end{document}